\documentclass{ws-ijmpe}

\begin{document}


%
%
\title{STUDY ON THE ENERGY DEPENDENCE OF THE RADII OF JETS
 BY THE HBT CORRELATION METHOD\\IN e$^+$e$^-$ COLLISIONS}

\author{\footnotesize CHEN ZHENG-YU, WANG MEI-JUAN, XIE YI-LONG, LIANG ZHU AND CHEN GANG$^\dagger$}

\address{Physical Department,School of Mathematics and Physics, China University of Geosciences\\
Wuhan, China, 430074\\
$^\dagger$chengang1@cug.edu.cn}

\maketitle


\begin{abstract}

The energy dependence of the radii size of jets are studied in
detail by the HBT correlation method using Monte Carlo Simulation
generator Jetset7.4 to produce 40,000,000 events of e$^+$e$^-$
collisions at $\sqrt s =30$, 50, 70, 91.2, 110, 130, 150 and 170
GeV. The radii of jets are measured using the HBT correlation method
with the indistinguishability of identical final state pions. It is
found that the average radii of quark-jets and gluon-jets are
independent of the c.m. energy of e$^+$e$^-$ collisions. The average
radius of quark-jets are obviously larger than that of gluon-jets.
The invariable average radii of quark-jets and gluon-jets in
e$^+$e$^-$ collisions are obtained at the end of parton evolvement.

\end{abstract}


\section{Introduction}

It was well-known that Hanbury-Brown and Twis had brought forward
HBT correlation in the process of measuring the angular radii of the
emitting sources in 1956\cite{use1}$^,$\cite{use2}$^,$\cite{use3}.
Later, the HBT correlation was widely used in subatom
studies\cite{use4}$^,$\cite{use5}$^,$\cite{use6}. The HBT
correlation method has been an important way to measure the size of
the emitting source in high energy collisions. Due to "color
confinement", we cannot observe free quarks and gluons and cannot
yet measure the size of them directly. However, the HBT correlation
method offers a viable indirectly method, and applying this method
into the high energy collisions we can obtain some characteristics
of strong interaction for quarks and gluons.

Historically, the discovery in 1975 of a two-jet
structure\cite{use7} in e$^+$e$^-$ collisions at center of mass
(c.m.) energy $\geq 6$ GeV had been taken as an experimental
confirmation of the parton model\cite{use71}, and the observation in
1979 of a third jet in
 e$^+$e$^-$ collisions at $17-30$ GeV had been recognized as the first experimental evidence
of the gluon \cite{use8,use81,use82,use83}. In the early 1990s, the
production of jets in hadron-hadron collisions was widely studied
 \cite{use84,use85,use86} and had been considered as an efficient way to obtain the
strong coupling constant $\alpha_s$ \cite{use87}. How to distinguish
jets and the study on jets are also very important, in relative high
energy ion collisions\cite{use9,use91}. Based on this idea we can
get information about quarks and strong interactions from the study
of jets by using the HBT correlation method.

In e$^+$e$^-$ collisions\cite{use10}, firstly, the e$^+$e$^-$ pair
is annihilated into a virtual $\gamma^*/Z^0$ resonance. The virtual
$\gamma^*/Z^0$, in turn, decays into a $q\bar{q}$ pair. Then the
initial $q\bar{q}$ may radiate other gluon and $q\bar{q}$ pairs,
giving rise to a cascade process. This stage is responsible for the
formation of hadronic jets. Further, the unstable hadrons decay into
experimentally observable particles (mostly pions). It has been
found that the majority of e$^+$e$^-$ collision events have a 2-jet
structure. If an initial quark or anti-quark emits a hard gluon with
sufficiently large transverse momentum, a 3-jet structure can be
formed. Thus, the source of a single jet is from a single initial
quark (or anti-quark) or gluon.

Although the quark and gluon, before being observed, have been
fragmented into the final state hadrons, the final state particles
inside the jets still carry a lot of information about the parent
quark and gluon. The quark and gluon are two different types of
particle. For example, the quark is a fermion with colour charge
equal to 4/3, while the gluon is a boson, carrying colour charge 3.
These differences will certainly influence their fragmentation,
resulting in different properties of quark-jets and gluon-jets.Some
characteristics of quarks (anti-quarks) or gluons is reflected by
the geometrical characteristics of jets. So, the study of the
geometrical characteristics of the jets is helpful in the
understanding of the perturbative/nonperturbative properties of QCD.

In the ref.\cite{use101}, the geometrical characters of quark-jets
and gluon-jet have been studied with the HBT correlation method
using MC generator producing quark-jets and gluon-jets in 3-jet
events of e$^+$e$^-$ collisions at $\sqrt s =91.2$GeV. However, do
the size of quark-jets and gluon-jets depend on the c.m. energy of
e$^+$e$^-$ collisions producing these jets? Are the size of
quark-jets measuring in 3- and 2-jets events of e$^+$e$^-$
collisions the same? Our work will focus on these questions.

 The paper is organized as follows: In Sec. II, we briefly
introduce the method of identification jets and the HBT correlation
function. In Sec. III, the average radius of quark-jets and
gluon-jets in 2-jet Events are calculated. In Sec. IV, the average
radius of quark-jets and gluon-jets in 3-jet Events are calculated.
A short summary is the content of Sec. V.

\section{The method of identification jets and HBT Correlation Function}

In our work, the data of e$^+$e$^-$ collision events are produced by
Monte Carlo Simulation generator Jetset7.4. The 2-jet events and the
3-jet events are selected using the Durham jet
algorithm\cite{use11}. In these methods, there is a cutting
parameter $y_{cut}$, which, in the case of the Durham algorithm, is
related to the relative transverse momentum $k_t$ as\cite{use111}
\begin{equation}
k_t=\sqrt{y_{cut}}\cdot \sqrt{s},
\end{equation}
 where $\sqrt s$ is the c.m. energy of the
collision. From the experimental point of view, $k_t$ can be taken
as the transition scale between the hard and soft processes. Its
value depends on the definition of "jet".

The single quark-jet and single gluon-jet are identified from 3-jet
events using the angular rule\cite{use12} .We assume that the three
jets in a 3-jet event come from quark, anti-quark and gluon,
respectively. Because of energy-momentum conservation, the three
jets in one event must lie in a plane, which is shown in Fig.1,
where $P_i(i=1,2,3)$ is the total momentum of all particles in
jet-$i$. The jets are tagged using the angles between them:

\begin{figure}[th]
\centerline{\psfig{file=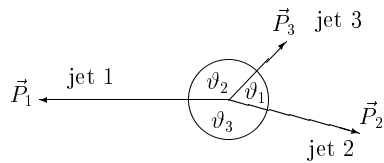,width=5cm}} \vspace*{4pt}
\caption{The sketch map of the 3-jet distribution.\label{f1}}
\end{figure}

\begin{equation}
\begin{array}{c}
\theta_i=\arccos\left(\frac{P_{j1}P_{k1}+P_{j2}P_{k2}+P_{j3}P_{k3}}
{\sqrt{P_{j1}^2+P_{j2}^2+P_{j3}^2}
\sqrt{P_{k1}^2+P_{k2}^2+P_{k3}^2}}\right),
\\
(i,j,k=1,2,3;i\neq j; j\neq k; k\neq i),
\end{array}
\end{equation}
where the largest angle, $\theta_3$, faces the gluon jet; the
smallest angle, $\theta_1$, faces the jet formed by an initial quark
without emitting a hard gluon; and the middle one, $\theta_2$, faces
the mother-quark-jet.

  According to the requirement of momentum conservation,
the three jets should be in one plane, and we add the condition:
$\theta_1+\theta_2+\theta_3>359^\circ$.

It should be noticed that the angle opposite to the mother-quark-jet
is very close to that opposite to the gluon-jet, i.e.
$\theta_2\approx \theta_3$, so they are easily confused. Therefore,
we demand a cut condition:  $\theta_3-\theta_2\geq \theta_{cut}$ ,
here $\theta_{cut}=10^\circ$. This cut rejects about 12$\%$
events\cite{use12}.


The HBT correlation, also called the Bose-Einstein correlation,
results from the indistinguishability of identical final state
particles. Most of the final state particles produced in $e^+e^-$
collisions are $\pi$ mesons, so we choose $\pi$ mesons
($\pi^+,\pi^-$, $\pi^0$) as the identical particles to study. If
$P(k_1,k_2)$ is defined as the probability of observing two
identical pions at the same time
 with momentum $k_1$ and $k_2$, and $P(k_1)$ and $P(k_2)$ are defined the
probability of observing pions with the momentum $k_1$ and $k_2$,
respectively. The correlation function $C_2(k_1,k_2)$ is defined as:
\begin{equation}
C_2(k_1,k_2)=\frac{P(k_1,k_2)}{P(k_1)P(k_2)},
\end{equation}

 If the equivalent density function of the source is parameterized
 to Gaussian form, we have:
\begin{equation}
\begin{array}{rcl}
C_2(k_1,k_2)&=&C_2(q,k_1,k_2)\\
&=&1+\lambda\exp\{-R_x^2q_x^2-R_y^2q_y^2-R_z^2q_z^2-\sigma_t^2q_t^2\},
\end{array}
\end{equation}
 where $q=k_1-k_2$ is the four-dimensional momentum difference.

If only the spatial part of the source is considered and assume that
the distribution of the source is isotropic, the correlation
function can be simplified as:
\begin{equation}
\begin{array}{lll}
C_2(k_1,k_2)&=&C_2(q,k_1,k_2)\\
&=&1+\alpha\exp\{-R^2Q^2\},
\end{array}
\end{equation}
where $\overrightarrow Q =\overrightarrow k_1-\overrightarrow k_2$
is the three-dimensional momentum difference. According to the
definition, jets do not possess spherical symmetry, but are axially
symmetric instead. So the parameter $R$ characterizing the
geometrical properties of the jets is actually the average radius of
jets.

In this paper, we study the average radius $R$ of the pion source
only through the spatial distribution function of the source, which
is taken as spherically symmetric. Then, the information about the
average size of the emitting source for the final state particles
can be obtained.

The three-dimensional momentum interval region chosen is
$Q=0\sim2.5$\ GeV/c, and is equally divided into 50 cells. We use
Monte Carlo simulation generator Jetset7.4 to produce e$^+$e$^-$
collision events both with and without HBT correlation, and then
select out suitable events for study. Identical $\pi$ mesons are
selected from the final state particles to make pion pairs after any
two $\pi$ mesons are grouped with each other. The three-dimensional
momentum difference of the $\pi$ meson pairs are calculated. The
correlation function (also called correlation coefficient) with
statistical method is:
\begin{equation}
C_2(k_1,k_2)=C_2(Q)=\frac{F_c(Q)}{F(Q)},
\end{equation}
where $F_c(Q)$ is the three-dimensional distribution function of the
identical particles with HBT correlation inside the jets and $F(Q)$
is the three-dimensional distribution function of the identical
particles without HBT correlation inside the jets. Since the
correlation among identical particles with large momentum difference
is quite weak, the distribution here with the HBT correlation should
be almost the same as the distribution without the HBT correlation.
Thus the $C_2(Q)$ can be multiplied by a coefficient to make the
value of it equal to 1. Thus, using Eq(5) to calculate the average
radius $R$, the Eq(6) can be expressed as:
\begin{equation}
C_2(Q)=\frac{F_c(Q)}{F(Q)}=\eta(1+\alpha\exp(-R^2Q^2)),
\end{equation}
where $\eta$ is the value of the correlation function $C_2(Q)$ at a
large momentum interval.

\section{The Measurement of source radii inside jets of 2-jet events}

We use Monte Carlo Simulation generator Jetset7.4 to produce
40,000,000 events of e$^+$e$^-$ collisions both with and without HBT
correlation, which the c.m. energies are $\sqrt s =30$, 50, 70,
91.2, 110, 130, 150, 170 GeV,respectively. The final state $\pi$
mesons ($\pi^+,\pi^-$, $\pi^0$) as the identical particles are
chosen from event samples to study. We just select out the final
state identical $\pi$ mesons emitted from vertex at origin, because
some of them are secondary emitted or even multistage emitted. Thus,
the calculated result is able to reflect the characteristics of the
source of jets properly. The 2-jet events are selected using the
Durham jet algorithm, and the cutting parameters at different c.m.
energies are selected as Table 1.

\begin{table}[th]
\tbl{The cutting parameters of selection 2-jet events from
e$^+$e$^-$ collision events. }
 {\begin{tabular}{@{}ccccccccc@{}}
 \toprule $\sqrt s$ (GeV)&30&50&70&91.2&110&130&150&170
 \\ \colrule
$y_{\rm cut}$&0.07&0.06&0.06&0.05&0.078&0.078&0.078&0.078\\
\botrule
\end{tabular} }
\end{table}

The 2-jet event is constituted by the two jets which are formed by
the fragmentation of the back-to-back $q\bar{q}$. And the two jets
are called quark jet-1 and quark jet-2, respectively. Due to the
back-to-back symmetry of the two jets formed by the fragmentation of
the $q\bar{q}$, the distribution patterns of the two jets should be
totally the same. So, we just need to study one of the two jets. We
will choose quark jet-1 which is referred to as quark-jet. The
correlation function is produced both with and without HBT
correlations when the three-dimensional momentum interval region of
the identical $\pi$ mesons is chosen as $Q=0\sim2.5$\ GeV. According
to equation (6) we calculate the value of the correlation function
$C_2(Q)$ of $\pi$ ($\pi^+,\pi^-$ and $\pi^0$) mesons inside
quark-jets from 2-jet events for all the 8 c.m. energies. Then, the
average radii size of emitting source of pion meson inside single
jet can be obtained through fitting the correlation functions
$C_2(Q)$ using Eq.(7) for $\pi^+,\pi^-$ and $\pi^0$ ,
 as shown in Table 2,respectively.
 As an example, the results for 3 c.m. energies are shown in Fig.2

\begin{figure}[th]
\centerline{\psfig{file=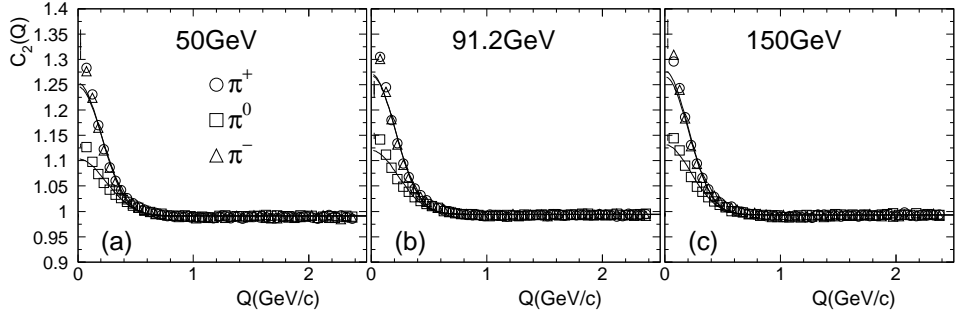,width=14cm}} \vspace*{9pt}
\caption{The distributions of correlation functions of quark-jet
from 2-jet events at different c.m. energies. The values of c.m.
energies are (a) 50 GeV, (b) 91.2GeV, (c) 150 GeV, respectively. The
data for the e$^+$e$^-$ collisions are produced by MC Jetset7.4
generator.\label{f1}}
\end{figure}

It is clear to see from Fig.2 that the distributions of correlation
functions of the $\pi$ meson inside quark-jets at different c.m.
energies are similar. Especially, the distributions of particle both
for $\pi^+$ and $\pi^-$ mesons at the same energies are in
superposition. However, the distributions of identical particles for
$\pi^0$ mesons is different to $\pi^+$ and $\pi^-$ mesons owing to
the electromagnetic interaction among $\pi^+$ and $\pi^-$ mesons in
the process of hadronization. The mean radii $R$ of jets for all
various c.m. energies are listed in Table 2 at the last line.

\begin{table}[th]
\tbl{The radii $R_{q}$ of quark-jets from 2-jet events at different
c.m. energies, measured by HBT using $\pi^+,\pi^0$ and $\pi^-$
mesons, respectively.}
 {\begin{tabular}{@{}cccc@{}}\colrule
 $\sqrt s$ (GeV)&$R_{q,\pi^+}$(fm)&$R_{q,\pi^0}$(fm)&$R_{q,\pi^-}$(fm)\\
\hline 30&0.704$\pm$0.006&0.553$\pm$0.007&0.708$\pm$0.006\\
50&0.662$\pm$0.004&0.580$\pm$0.006&0.686$\pm$0.004\\
70&0.703$\pm$0.004&0.577$\pm$0.005&0.688$\pm$0.004\\
91.2&0.672$\pm$0.003&0.607$\pm$0.006&0.684$\pm$0.004\\
110&0.697$\pm$0.003&0.593$\pm$0.004&0.678$\pm$0.004\\
130&0.710$\pm$0.004&0.616$\pm$0.005&0.679$\pm$0.003\\
150&0.693$\pm$0.005&0.624$\pm$0.007&0.685$\pm$0.005\\
170&0.705$\pm$0.004&0.650$\pm$0.008&0.715$\pm$0.004\\
\hline $R_{\rm aver}$&0.693$\pm$0.004&0.600$\pm$0.006&0.690$\pm$0.004\\
\botrule
\end{tabular} }
\end{table}

\begin{figure}[th]
\centerline{\psfig{file=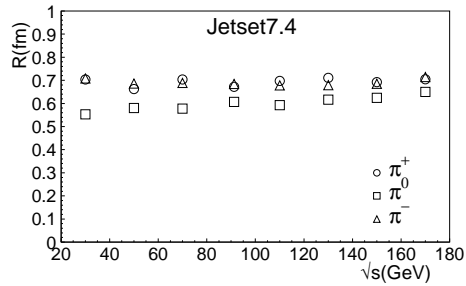,width=6cm}} \vspace*{1pt}
\caption{The comparison of the values of the radii $R$ of quark-jets
from 2-jet events for different types of $\pi$ mesons
($\pi^+,\pi^-$, $\pi^0$) at different c.m. energies.\label{f1}}
\end{figure}

For the convenience of comparison, we draw the source radii of the
three kinds of $\pi$ mesons inside quark-jets for all energies in
Fig.3.

 It can be seen from Table 2 or Fig.3 that the values of the
radii $R$ of jets for one meson from different c.m. energies are
nearly the same within the error range; the radii $R$ of jets for
the same c.m. energy both $\pi^+$ and $\pi^-$ mesons are
approximately the same within the error range, i.e. their means
$\bar{R}_{q,\pi^+}=0.693\pm 0.003\ \rm{fm}$,
$\bar{R}_{q,\pi^-}=0.690\pm 0.004\ \rm{fm}$; But there are some
difference between the $\pi^+$ or $\pi^-$ mesons and $\pi^0$ mesons
for the $\bar{R}_{q,\pi^0}=0.6000\pm 0.006\ \rm{fm}$. This is due to
the electromagnetic interaction among $\pi^+$ and $\pi^-$ mesons in
the process of hadronization. So, the results for $\pi^0$ mesons are
more authentic.

\section{The Measurement of source radii inside jets of 3-jet events}

In the same way, we use Monte Carlo Simulation generator Jetset7.4
to produce 40,000,000 events of e$^+$e$^-$ collisions both with and
without HBT correlation for 8 c.m. energies. And we choose the final
state $\pi$ mesons ($\pi^+,\pi^-$, $\pi^0$) emitted from the vertex
at origin as the identical particles to study. The 3-jet events are
selected using the Durham jet algorithm, which the cutting
parameters for 8 energies are listed in Table 3.

\begin{table}[th]
 \tbl{The cutting parameters of selection 3-jet events from e$^+$e$^-$ collision events. }
 {\begin{tabular}{@{}ccccccccc@{}} \toprule $\sqrt s$ (GeV)&30&50&70&91.2&110&130&150&170
 \\ \colrule
$y_{\rm cut}$&0.008&0.005&0.004&0.002&0.0015&0.001&0.0008&0.0005\\
\botrule
\end{tabular} }
\end{table}

After the quark-jet, mother quark-jet and gluon-jet are identified
from 3-jet events using the angular rule, the correlation function
is produced both for the case with and without HBT correlation when
the three-dimensional momentum interval region of the identical
$\pi$ mesons is chosen as $Q=0\sim2.5$\ GeV/c. We calculate the
value of $C_2(Q)$ according to formula (6), then the average radius
size of emitting source of pion inside single jet can be obtained
through fitting the correlation functions $C_2(Q)$ using Eq.(7) with
$\pi^+,\pi^-$ and $\pi^0$ meson, respectively. The results are
listed in Table 4. As an example, the results for 3 c.m. energies
are shown in Fig.4.

\begin{figure}[th]
\centerline{\psfig{file=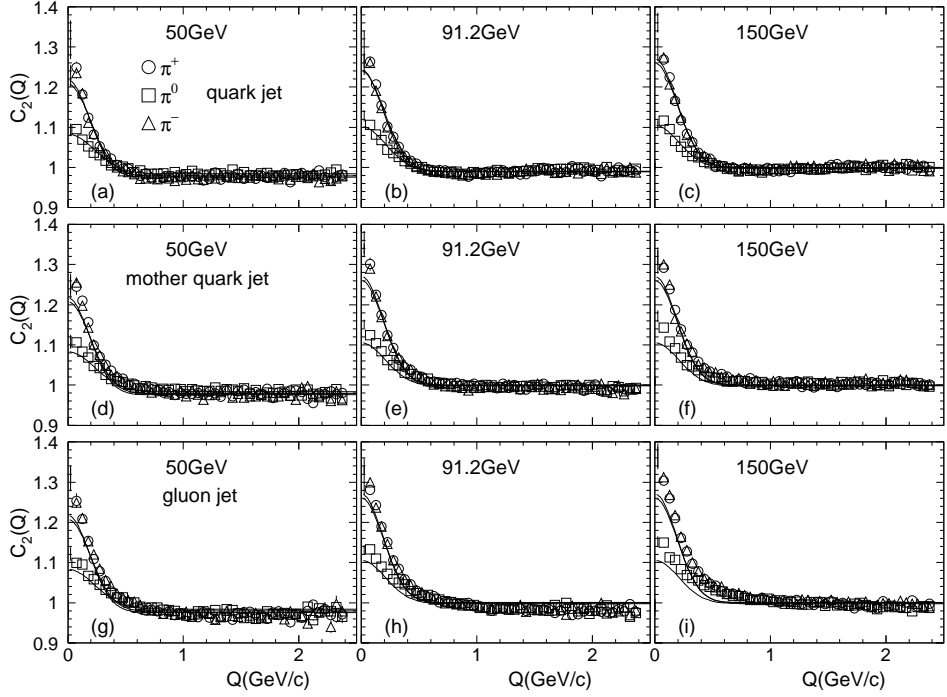,width=14cm}} \vspace*{9pt}
\caption{The distributions of $\pi$ meson correlation functions of
the three kinds of jets from 3-jet events at different c.m.
energies. The first row is for the case of  quark-jets; the second
row is for the case of mother quark-jets; the third line row is for
the case of gluon-jets. The c.m. energies of (a), (d) and (g) are 50
GeV; (b), (e) and (h) are 91.2 GeV; (c), (f) and (i) are 150
GeV,respectively. The data for the e$^+$e$^-$ collisions are
produced by MC Jetset7.4 generator.\label{f1}}
\end{figure}

It is easy to come to the conclusion from Fig.4 that the three type
of $\pi$ meson correlation functions for quark-jets, mother
quark-jets and gluon-jets are all similar. And the distributions of
different identical particles, for $\pi^+$ and $\pi^-$ mesons,
formed at the same energy are nearly in superposition. However, the
distributions of identical particles for $\pi^0$ mesons are
different to $\pi^+$ and $\pi^-$ mesons.

\begin{table}[th]
\tbl{The values of the source radii $R$ of the three types of jets
at different c.m. energies calculated with the three kinds of $\pi$
mesons ($\pi^+,\pi^-$, $\pi^0$). }
 {\begin{tabular}{@{}ccccc@{}} \toprule $\pi$ mesons&$\sqrt s$
 (GeV)&$R_{\rm quark}$(fm)&$R_{\rm m-quark}$(fm)&$R_{\rm gluon}$(fm)
 \\ \colrule
&30&0.60$\pm$0.03&0.63$\pm$0.02&0.69$\pm$0.02\\
&50&0.72$\pm$0.02&0.64$\pm$0.02&0.66$\pm$0.01\\
&70&0.73$\pm$0.01&0.70$\pm$0.01&0.54$\pm$0.01\\
$\pi^+$&91.2&0.71$\pm$0.01&0.68$\pm$0.01&0.64$\pm$0.01\\
&110&0.74$\pm$0.01&0.64$\pm$0.01&0.55$\pm$0.01\\
&130&0.74$\pm$0.01&0.66$\pm$0.01&0.56$\pm$0.01\\
&150&0.73$\pm$0.01&0.68$\pm$0.01&0.56$\pm$0.01\\
&170&0.75$\pm$0.01&0.70$\pm$0.02&0.55$\pm$0.01\\
&$R_{\rm aver}$&0.71$\pm$0.02&0.67$\pm$0.01&0.59$\pm$0.01\\
\hline &30&0.44$\pm$0.02&0.41$\pm$0.02&0.48$\pm$0.02\\
&50&0.60$\pm$0.01&0.54$\pm$0.02&0.46$\pm$0.01\\
&70&0.61$\pm$0.01&0.53$\pm$0.01&0.42$\pm$0.01\\
$\pi^0$&91.2&0.67$\pm$0.02&0.53$\pm$0.01&0.44$\pm$0.01\\
&110&0.64$\pm$0.02&0.60$\pm$0.01&0.42$\pm$0.01\\
&130&0.60$\pm$0.01&0.57$\pm$0.02&0.41$\pm$0.01\\
&150&0.67$\pm$0.02&0.61$\pm$0.02&0.41$\pm$0.01\\
&170&0.66$\pm$0.02&0.59$\pm$0.02&0.44$\pm$0.01\\
&$R_{\rm aver}$&0.61$\pm$0.02&0.55$\pm$0.02&0.43$\pm$0.01\\
\hline &30&0.76$\pm$0.02&0.65$\pm$0.03&0.67$\pm$0.02\\
&50&0.75$\pm$0.02&0.66$\pm$0.01&0.62$\pm$0.01\\
&70&0.72$\pm$0.01&0.70$\pm$0.02&0.60$\pm$0.01\\
$\pi^-$&91.2&0.75$\pm$0.01&0.65$\pm$0.02&0.61$\pm$0.01\\
&110&0.73$\pm$0.01&0.67$\pm$0.01&0.55$\pm$0.01\\
&130&0.75$\pm$0.01&0.69$\pm$0.01&0.57$\pm$0.01\\
&150&0.74$\pm$0.01&0.67$\pm$0.01&0.60$\pm$0.01\\
&170&0.72$\pm$0.01&0.72$\pm$0.02&0.59$\pm$0.01\\
&$R_{\rm aver}$&0.74$\pm$0.01&0.68$\pm$0.02&0.60$\pm$0.01\\
\hline \botrule
\end{tabular} }
\end{table}

For the convenient of comparison, we draw the average source radii
of the three kinds of jets at different c.m. energies for the three
types of $\pi$ mesons in Fig.5.

\begin{figure}[th]
\centerline{\psfig{file=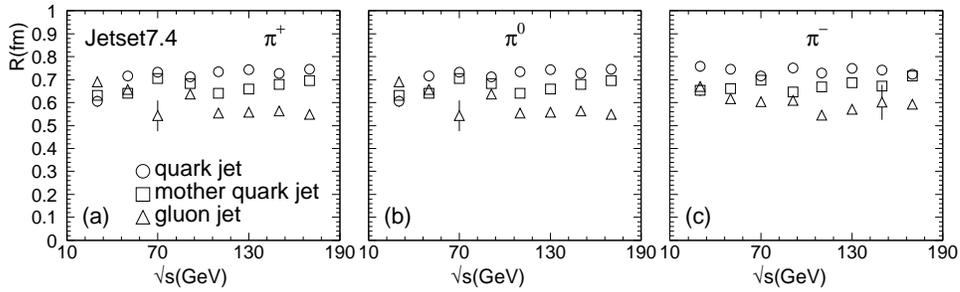,width=14cm}} \vspace*{9pt}
\caption{The comparison of the values of the radii $R$ of the three
kinds of jets from 3-jet events at different c.m. energies for
different $\pi$ mesons, (a) $\pi^+$, (b) $\pi^0$, (c)
$\pi^-$.\label{f1}}
\end{figure}

It can be seen from Table 4 and Fig.5 that: the values of radii $R$
of quark-jets or mother quark-jets and or gluon-jets at different
c.m. energies are all nearly the same within the error range,
respectively; the average radii of quark-jets are obviously larger
than that of gluon-jets, and the average radii of quark-jets are
also larger than that of mother quark-jets. The mean radius of
quark-jets for $\pi^+$ mesons is $0.71\pm 0.02\ \rm{fm}$, that of
mother quark-jets is $0.67\pm 0.01\ \rm{fm}$, and that of gluon-jets
is $0.59\pm 0.01\ \rm{fm}$. The mean radius of quark-jets for
$\pi^0$ mesons is $0.61\pm 0.02\ \rm{fm}$, that of mother quark-jets
is $0.55\pm 0.02\ \rm{fm}$, and that of gluon-jets is $0.43\pm 0.01\
\rm{fm}$. The mean radius of quark-jets for $\pi^-$ mesons is
$0.74\pm 0.01\ \rm{fm}$, that of mother quark-jets is $0.68\pm 0.02\
\rm{fm}$, and that of gluon-jets is $0.60\pm 0.01\
\rm{fm}$,respectively.

\section{Conclusion and discussion}

We use Monte Carlo Simulation generator Jetset7.4 to produce the
data of e$^+$e$^-$ collision events, which the c.m. energies are
$\sqrt s = 30$, 50, 70, 91.2, 110, 130, 150, 170 GeV. The 2-jet
events and 3-jet events are selected using the Durham jet algorithm.
The geometrical characters of quark-jets and gluon-jets are
 studied in detail using the HBT correlation method. The conclusions
are as follows:

\begin{itemlist}

\item  The radii of quark-jets or gluon-jets
measured at different c.m. energies are approximately the same
within the error range, which shows that the radii of quark-jets and
gluon-jets reflect some intrinsic properties of quarks and gluons.

\item The values of the mean radii of quark-jets, mother quark-jets and
gluon-jets are measured by calculating the final state identical
$\pi$ mesons using the HBT correlation method are obtained, shown in
Table 5.
\begin{table}[th]
\tbl{The means radii $R$ of jets measured using the HBT correlation
method in 2-jet events and 3-jet events from e$^+$e$^-$ collision
events. }
  {\begin{tabular}{@{}ccccc@{}} \toprule  $\pi$ mesons&2-jet events&&3-jet events& \\ \colrule
&$R_{\rm quark}$(fm)&$R_{\rm quark}$(fm)&$R_{\rm
m-quark}$(fm)&$R_{\rm gluon}$(fm)
\\ \colrule
$\pi^+$&0.693$\pm$0.004&0.71$\pm$0.02&0.67$\pm$0.01&0.59$\pm$0.01\\
$\pi^0$&0.600$\pm$0.006&0.61$\pm$0.02&0.55$\pm$0.02&0.43$\pm$0.01\\
$\pi^-$&0.690$\pm$0.004&0.74$\pm$0.01&0.68$\pm$0.02&0.60$\pm$0.01\\
\botrule
\end{tabular} }
\end{table}

\item The mean radii of quark-jets measured is
obviously larger than that of gluon, which indicates that the
 size of quark is larger than that of gluon. However, the mean
  radii of mother quark-jets measured is less than that
  of quark-jets. This may be due to the mixture of a small amount of
  gluon-jets and mother-quark-jets in the process of
  measurement which makes the radii of mother quark-jets measured smaller.

  \item The results for $\pi^0$ mesons are more authentic than for $\pi^+$
and $\pi^-$ mesons for there are no electromagnetic interactions
among $\pi^0$ mesons in the process of hadronization.
\end{itemlist}

\vspace{-2mm} \centerline{\rule{80mm}{0.1pt}} \vspace{2mm}
\section*{Acknowledgments}
This work is supported by self-determined and innovative research
funds of CUG(1210491B10) and Research Funds for Central Universities
(GUGL 100237) ¡£

\end{document}